\documentclass[12pt,english]{article}
\usepackage{babel}
\usepackage{amsmath}
\usepackage{amssymb} 
\usepackage{bbm} 
\usepackage[small]{caption2} 
\usepackage{fleqn} 
\usepackage{graphicx} 
\usepackage[small,loose]{subfigure}  
\usepackage{cite} 

\advance \headheight by 3.0truept       
\setcaptionwidth{.85\textwidth}

\DeclareMathOperator{\diag}{diag}

\newcommand{\CenterObject}[1]{\ensuremath{\vcenter{\hbox{#1}}}}

\newcommand{\I}{\mathrm{i}}

\def\p{\pi}

\def\s{\sigma}

\def\J{\Psi}

\def\beq{\begin{equation}}
\def\eeq{\end{equation}}
\def\bea{\begin{eqnarray}}
\def\eea{\end{eqnarray}}
\newcommand{\E}[1]{\ensuremath{\mathrm{E}_{#1}}} 
\newcommand{\G}[1]{\ensuremath{\mathrm{G}_{#1}}}
\newcommand{\SO}[1]{\ensuremath{\mathrm{SO}(#1)}}
\newcommand{\SU}[1]{\ensuremath{\mathrm{SU}(#1)}}
\newcommand{\U}[1]{\ensuremath{\mathrm{U}(#1)}}
\newcommand{\Z}[1]{\ensuremath{\mathbbm{Z}_{#1}}} 
\def\zc{\Z{6}} 
\def\zn{\Z{N}} 
\def\za{\Z{3}} 
\def\zb{\Z{2}} 
\def\z{\Z{}} 

\def\pl#1#2#3{Phys.~Lett.~{\bf B {#1}} ({#2}) #3}
\def\np#1#2#3{Nucl.~Phys.~{\bf B {#1}} ({#2}) #3}
\def\prl#1#2#3{Phys.~Rev.~Lett.~{\bf #1} ({#2}) #3}
\def\pr#1#2#3{Phys.~Rev.~{\bf D {#1}} ({#2}) #3}

\def\ap#1#2#3{Ann.~of Phys.~{\bf {#1}} ({#2}) #3}

\def\ptp#1#2#3{Progr.~Theor.~Phys.~{\bf {#1}} ({#2}) #3}

\renewcommand{\(}{\left(}
\renewcommand{\)}{\right)}
\renewcommand{\[}{\left[}

\begin{document}
\title{
\begin{flushleft}
{\normalsize DESY 04-237\hfill December 2004}
\end{flushleft}
\vspace{2cm}
{\bf Dual Models of Gauge Unification\\ in Various  Dimensions}\\[0.8cm]}
\author{\\
{\bf\normalsize 
Wilfried Buchm\"uller, Koichi Hamaguchi, Oleg Lebedev} \\[0.2cm] 
{\it\normalsize Deutsches Elektronen-Synchrotron DESY, 22603 Hamburg, Germany}
\\[0.4cm] 
{\bf\normalsize Michael Ratz}\\[0.2cm]
{\it\normalsize Physikalisches Institut der Universit\"at Bonn,}\\
{\it\normalsize Nussallee 12, 53115 Bonn, Germany}
 }
\date{}
\maketitle \thispagestyle{empty} 

\abstract{
\noindent
We construct a compactification of the heterotic string on an orbifold 
$T^6/\Z6$
leading to the standard model spectrum plus vector--like  matter. The standard
model gauge group is obtained as an intersection of three $\SO{10}$ subgroups 
of
$\E{8}$.  Three families of $\SO{10}$  ${\bf 16}$-plets are localized at three
equivalent fixed points. Gauge coupling unification favours existence of  an
intermediate GUT which can have any dimension between five and ten. 
Various GUT gauge groups occur. For example, in six dimensions one can have
$\E{6}\times \SU3$,  $\SU4\times\SU4 \times \U1^2$ or  $\SO{8}\times \SO{8}$, 
depending on which of the compact dimensions are large. The different
higher--dimensional  GUTs are `dual' to each other. They represent different
points in moduli space, with the same massless spectrum and ultraviolet 
completion.}

\clearpage

\section{Embedding the standard model in $\boldsymbol{\E8}$}
\label{sec:SMinE8}

The symmetries and the particle content of the standard model point towards 
grand unified theories (GUTs) as the next step in the unification of all 
forces. Left- and right-handed quarks and leptons
can be grouped in three \SU5 multiplets \cite{gg74},  
$\boldsymbol{10} = (q_\mathrm{L},u_\mathrm{R}^c,e_\mathrm{R}^c)$, 
$\boldsymbol{\overline{5}} = (d_\mathrm{R}^c,\ell_\mathrm{L})$ and 
$\boldsymbol{1}= \nu_\mathrm{R}^c$. Here we have added right-handed neutrinos which are
suggested by the evidence for neutrino masses.
All quarks and leptons of one generation can be unified in a single multiplet 
of the GUT group \SO{10} \cite{gfm75},
\begin{equation}
\boldsymbol{16} = \boldsymbol{10} + \boldsymbol{\overline{5}} +\boldsymbol{1}\;. 
\end{equation}
The group \SO{10} contains as subgroups the Pati-Salam group \cite{ps74}, 
$G_{\rm PS} = \SU4 \times \SU2 \times \SU2$, the Georgi-Glashow 
group $\SU5$, 
$G_{\rm GG} = \SU5 \times \U1$, and the `flipped' $\SU5$ group, 
$G_{\rm fl} = \SU5' \times \U1'$
\cite{bar82}, where the right-handed up- and down-quarks are interchanged, 
yielding another viable GUT group.

It is a remarkable property of the standard model that the matter fields form
complete $\SO{10}$ multiplets whereas the gauge and Higgs fields are 
`split multiplets'. They have to be combined with other split multiplets,
not contained in the standard model, in order to obtain a complete unified
theory. It is also well known \cite{oli81} that exceptional groups play an
exceptional role in grand unification, and the embedding
\begin{equation}\label{emb}
\SO{10} \subset \E6 \subset \E8
\end{equation}
appears, in particular, in  compactifications of the 
heterotic string \cite{ghx85}  on Calabi-Yau manifolds \cite{chx85}.

As we shall see, complete $\SO{10}$ matter multiplets together with split
gauge and Higgs multiplets arise naturally in orbifold compactifications
of higher-dimensional unified theories. Orbifold compactifications 
have first been considered in string theory \cite{dhx85,inx87} and 
subsequently
in effective higher-dimensional field theories \cite{kaw00,abc01}. They
provide a simple and elegant way to break GUT symmetries, while
avoiding the notorious doublet-triplet splitting problem. 
More recently, it has been shown how orbifold GUTs can occur in orbifold
string compactifications \cite{krz041,fnx04,krz042}.

In the following we shall first search for a scheme of $\zn$ twists which allows
to break $\E8$, a common ingredient of string models, to the standard model
group. A $\zn$ twist is an element of the gauge group $G$, with
\begin{equation}
P = \exp{(-2\pi \I V_N\cdot \boldsymbol{H})}\;,\quad P^N = \mathbbm{1}\;.
\end{equation}
Here the generators $\boldsymbol{H}_i$ form the (Abelian) Cartan
subalgebra of $G$, and  $V_N$ is a real vector. 
The twist $P$ acts on the Cartan and step generators 
$\boldsymbol{E}_\alpha$ as follows:
\begin{eqnarray}
P\,\boldsymbol{H}_i\,P^{-1} & = & \boldsymbol{H}_i \, \nonumber\\
P\,\boldsymbol{E}_\alpha\,P^{-1} & = & 
\exp{(-2\pi \I V_N \cdot \alpha)}\, \boldsymbol{E}_\alpha \;,
\end{eqnarray}
where $\alpha$ is a root associated with $\boldsymbol{E}_\alpha$.
Clearly, $P$ 
breaks $G$ to a subgroup containing all step generators 
which commute with $P$, i.e. $[P,\boldsymbol{E}_\alpha]=0$.

\begin{figure}[t]
 \centerline{\CenterObject{\includegraphics{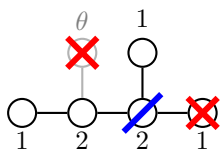}}}
 \caption{$\SO{10}$ breaking patterns by a $\Z2$ twist.
 The action of the Pati-Salam twist is indicated by crosses, while that of
 the Georgi-Glashow twist is indicated by  a slash.}
 \label{fig:SMinSO10}
\end{figure}
The symmetry breaking is conveniently expressed in terms of the  
 Dynkin diagrams. This technique has been employed 
to classify possible symmetry breaking patterns in string models
\cite{Breit:1985ud,kkx90,Choi:2003pq}
and, more recently, in  orbifold GUTs \cite{hmr02,hr03}.
Starting with the extended Dynkin diagram which contains
the most negative root in addition to the simple roots,
regular subgroups of a given group are obtained by crossing
out some of the roots of the Dynkin diagram.
In particular, the action of the $\zn$ orbifold twist 
essentially amounts to crossing out a root with the (Coxeter) 
label $N$, or more generally,  roots whose labels sum up to  $\leq N$.

As an example, consider the breaking of $\SO{10}$, displayed in
Fig.~\ref{fig:SMinSO10}. For each simple root the Coxeter label is listed, which
determines the order of the twist required for the corresponding symmetry
breaking. Crossing out one of the nodes with label 2 breaks $\SO{10}$ to the
semi-simple subgroup $G_{\rm PS}$, while crossing out  one of the roots with
label 1 together with the most negative root $\theta$ breaks  $\SO{10}$   to
$G_{\rm GG}$. The intersection of the two groups gives the standard model with
an additional $\U1$ factor \cite{abc01},
\begin{equation}
G_{\rm GG} \cap G_{\rm PS} = \SU3 \times \SU2 \times \U1^2 \sim G_{\rm SM} \;,
\end{equation}
where `$\sim$' means `modulo $\U1$ factors'.
Under the $\Z2$ twisting, 
the group generators divide into those with positive and negative 
parities  $P$ with respect to the twist. 
Combining the two parities $P_{\rm GG}$ and
$P_{\rm PS}$, one can construct the third $\Z2$ parity
$P_{\rm GG}\cdot P_{\rm PS} = P_{\rm fl}$ 
which breaks $\SO{10}$ to the flipped $\SU5$,
\begin{equation}
\SO{10} \stackrel{P_{\rm fl}}{\longrightarrow} G_{\rm fl} = \SU{5}'
\times \U{1}'\;,
\end{equation} 
The standard model group $G_{\rm SM}$ can also be obtained as an 
intersection of
the two $\SU5$ embeddings, $G_{\rm GG}$ and $G_{\rm fl}$,
\begin{equation}
G_{\rm GG} \cap G_{\rm fl} = \SU3 \times \SU2 \times \U1^2 \sim G_{\rm SM} \;.
\end{equation}

\begin{figure}[t]
 \centerline{\CenterObject{\includegraphics{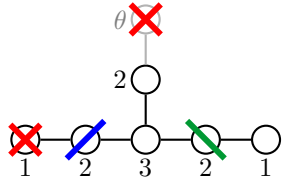}}}
 \caption{$\E6$ breaking patterns under  $\Z2$ twisting. Three different $\Z2$ twists
are indicated by crosses, a slash and a backslash, respectively.}
 \label{fig:SMinE6}
\end{figure}

As another example, consider now  $\E6$ breaking to the standard model group.
From the extended Dynkin diagram Fig.~\ref{fig:SMinE6} it is clear, in analogy
with the  $\SO{10}$ breaking, that three $\zb$ twists,
\begin{eqnarray}
\E6 & \stackrel{P_A}{\longrightarrow} &\SO{10} \times \U1\;, \quad
\E6 \,\stackrel{P_B}{\longrightarrow}\, \SU6'\times \SU2' \;, \nonumber\\
\E6 & \stackrel{P_C}{\longrightarrow} & \SU6''\times \SU2'' \;,
\end{eqnarray}
can break $\E6$ to the standard model up to $\U1$ factors,
\begin{multline}
\SO{10} \times \U1\ \cap\ \SU6'\times \SU2'\ \cap\ \SU6''\times \SU2''
\\
{}= \ \SU3 \times \SU2 \times \U1^2\ \sim \ G_\mathrm{SM}\;.
\end{multline}
\begin{figure}[t]
 \centerline{\CenterObject{\includegraphics{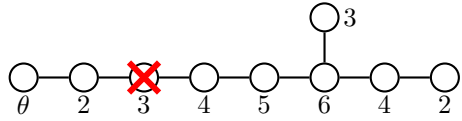}}}
 \caption{$\E8$ breaking to $\E6\times \SU3$.}
 \label{fig:E6inE8}
\end{figure}
As in the $\SO{10}$ example, one can check that
the same breaking can be obtained as an  intersection of three 
different $\SO{10}$ embeddings  in $\E6$ which correspond to the twists 
\begin{equation}
\E6 \xrightarrow{P_A\cdot P_C} \SO{10}' \times \U1'\;, \quad
\E6 \xrightarrow{P_A\cdot P_B\cdot P_C} \SO{10}'' \times \U1''\;,
\end{equation}
such that
\begin{multline}
\SO{10}\times \U1\ \cap\ \SO{10}'\times \U1'\ \cap\ \SO{10}''\times \U1''\\
=\ \SU3 \times \SU2 \times \U1^2\ \sim \ G_\mathrm{SM}\;.
\end{multline}
Let us remark that it is not possible to distinguish the three
$\SO{10}$ embeddings in $\E6$ (as well as $G_\mathrm{GG}$ and
$G_\mathrm{fl}$ embeddings in $\SO{10}$) 
at the level of Dynkin diagrams. 
The corresponding
subalgebras are related by Weyl reflections within the embedding group.
To distinguish them, an explicit analysis of the shift vectors $V_N$  is required.

Our final goal is to break $\E8$  to the standard model gauge group. This can be
achieved by combining the above three $\zb$ twists with a $\za$ twist which
breaks $\E8$ to $\E6\times \SU3$ (cf.~Fig.~\ref{fig:E6inE8}). The $\zb$ twists can then also
break the $\SU3$ factor to $\SU2\times \U1$. In this way one obtains three $\zc$
twists which break $\E8$ to subgroups containing $\SO{10}$, 
\begin{eqnarray}
\E8 &\stackrel{P_6}{\longrightarrow} & \SO{10} \times \SU3 \times \U1\;, \quad
\E8 \,\stackrel{P_6'}{\longrightarrow}\, \SO{10}'\times \SU2'\times \U1^2\;,
\nonumber\\
 \E8 &\stackrel{P_6''}{\longrightarrow} & 
 \SO{10}''\times \SU2'' \times \U1^2\;, 
\end{eqnarray}
such that the intersection is the standard model group up to $\U1$ factors,
\begin{multline}
\SO{10} \times \SU3 \times \U1\ \cap\ \SO{10}'\times \SU2'\times \U1^2\\
\cap\ \SO{10}''\times \SU2'' \times \U1^2\
\sim\ G_\mathrm{SM}\;.
\end{multline}
In an orthonormal basis of $\E8$ roots,
three $\zc$ shift  vectors which realize the described symmetry breaking
read explicitly:
\begin{eqnarray}
V_6 &=& \left({1\over 3},{1\over 3},{1\over 3},0,0,0,0,0\right)\;, \nonumber \\
V_6'&=& \left({7\over 12},{7\over 12},{1\over 12},{1\over 4},-{1\over 4},
              -{3\over 4},-{3\over 4},-{3\over 4}\right)\;, \nonumber\\
V_6'' &=& \left({7\over 12},{13\over 12},{7\over 12},{3\over 4},{1\over 4},
              -{3\over 4},-{3\over 4},-{3\over 4}\right)\;.
\label{3V6}
\end{eqnarray}
Note that the differences between the $\zc$ shift vectors 
are $\zb$ shift vectors,
\begin{eqnarray}
W_2 &=& V_6''-V_6' = \left(0,{1\over 2},{1\over 2}, {1\over 2},{1\over 2},
                       0,0,0 \right)\;, \nonumber \\
W_2' &=& V_6'-V_6 = \left( {1\over 4}, {1\over 4},- {1\over 4},{1\over 4},
-{1\over 4},-{3\over 4},-{3\over 4},-{3\over 4} \right)\;,
\end{eqnarray}
which will play the role of Wilson lines in the next section.

To summarize, in this section we have
presented a group--theoretical analysis of $\E8$
breaking to the standard model with intermediate $\E6$ and $\SO{10}$ GUTs,
suggested by the structure of matter multiplets.

\section{Orbifold compactification}

Let us now construct an orbifold compactification of the heterotic string, 
which realizes the symmetry breaking described above. As is clear
from the above discussion, we will need a
$\zc$ or a higher--order orbifold and choose the former for simplicity. 

In the light cone gauge the heterotic string \cite{ghx85} can be described
by the following bosonic world sheet fields: 8 string coordinates
$X^i$, $i= 1\ldots 8$, 16 internal left-moving coordinates $X^I$,
$I=1\ldots 16$, and 4 right-moving fields $\phi^i$, $i=1\ldots 4$, which
correspond to the bosonized Neveu-Schwarz-Ramond fermions 
(cf.~\cite{imx88,kkx90,bl99}). The 16 left-moving internal coordinates
are compactified on a torus. The associated quantized momenta lie on the
$\E8\times \E8$ root lattice. In an orthonormal basis, vectors of the
$\E8$ root lattice are given by
\begin{equation}
p_{_{\E8}}= (n_1,...,n_8)\;\;{\rm or}\;\; 
\(n_1 +{1\over 2}, ..., n_8 +{1\over 2} \)\;,
\label{root}
\end{equation}
with integer $n_i$ satisfying $\sum_{i=1}^8n_i=0 \mod 2$.
The massless spectrum of this 10D string is 10D supergravity coupled to 
$\E8\times \E8$ super Yang--Mills theory.

To obtain a four--dimensional theory, 6 dimensions of the 10D heterotic string
are compactified on an orbifold. In our case, this is a $\zc$ orbifold
obtained by modding a 6D torus together with   the 16D gauge torus
by a $\zc$ twist,
\begin{equation}
{\cal O}\ =\ T^6 \otimes T_{\E8\times \E8'} / \zc\;.
\end{equation}
On the three complex torus coordinates $z^i$, $i=1,2,3$, the $\zc$ twist 
acts as 
\begin{equation}
z^i \rightarrow e^{2\pi \I\, v_6^i}~ z^i\;.  
\label{v6}
\end{equation}
Here $6v_6$ has integer components. The compact string coordinates are
described by the complex variables $Z^i=X^{2i-1}+\I\ X^{2i}$,
$i=1\ldots 3$. The $\zc$ action on the string
coordinates reads, up to lattice translations (cf.~\cite{kkx90}), 
\begin{equation}
Z^i(\s=2\p) = e^{2\p\I k v_6^i}\ Z^i(\s=0)\;,\quad k=0\ldots 5\;,
\end{equation}
\begin{equation}
\phi^i(\s=2\p) = \phi^i(\s=0) - \p k v_6^i\;, \quad
X^I(\s=2\p) = X^I(\s=0) + \p k V_6^I\;, 
\end{equation}
where $6V^I$ is an $\E8\times \E8$ lattice vector. 

The torus $T^6$ is spanned by basis vectors $e_\kappa$, $\kappa=1,...,6$.
In general, a torus allows for the presence of Wilson lines,
i.e. a translation by a lattice vector $n_\kappa e_\kappa^i$ can be
accompanied by a shift of the internal string coordinates,
\begin{eqnarray}
X^i(\s=2\p) &=& X^i(\s=0) + 2\p n_\kappa e_\kappa^i\;,\quad  
n_\kappa \in \z\;,\nonumber\\
X^I(\s=2\p) &=& X^I(\s=0) + \p n_\kappa W_\kappa^I\;.
\end{eqnarray}
Here the discrete Wilson lines $W_\kappa$ are restricted by symmetry and by
modular invariance.

The basis vectors $e_\kappa$ 
are taken to be simple roots  of a Lie algebra, whose choice is dictated by
the required symmetry of the lattice.
In our case the  lattice must have a $\zc$ symmetry and allow for
the existence of 3 independent $V_6$ shift vectors (\ref{3V6}) 
(or two  Wilson lines of order 2). 
This leaves two possibilities for the Lie lattice \cite{kkx90}:
\begin{equation}
\G2 \times \SU3 \times \SO4\;\;{\rm or}\;\; \SU3^{[2]}\times \SU3\times \SO4\;.
\end{equation}
We shall base our analysis on the first lattice, which has recently been
studied in detail by Kobayashi, Raby and Zhang \cite{krz041}. These authors
have obtained models with the Pati-Salam gauge group in four dimensions, which then has 
to be broken to the standard model by the Higgs mechanism. The model described
in the following differs from those in the choice of $\zc$ twists and the pattern
of symmetry breaking.

\begin{figure}[t]
 \centerline{\begin{tabular}{lccc}
  & $\mathrm{G}_2$ torus & $\SU3$ torus & $\SO4$ torus\\
  $ \zc~;~T_{1,5}$ & 
  \CenterObject{\includegraphics[scale=0.8]{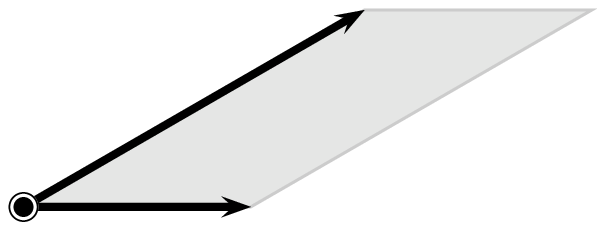}}
  &
  \CenterObject{\includegraphics[scale=0.8]{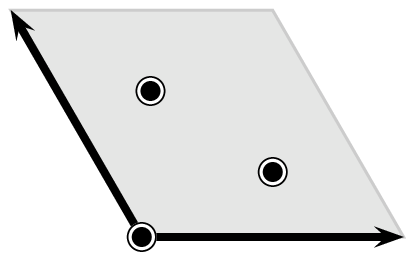}}
  &
  \CenterObject{\includegraphics[scale=0.8]{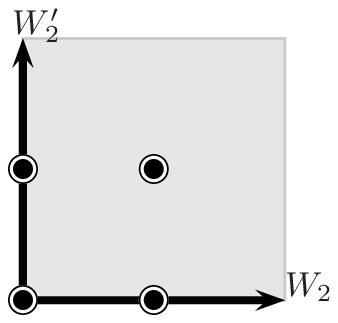}}
  \\[2cm]
  $\za~;~T_{2,4}$ &
  \CenterObject{\includegraphics[scale=0.8]{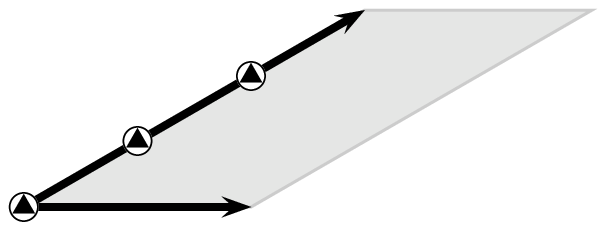}}
  &
  \CenterObject{\includegraphics[scale=0.8]{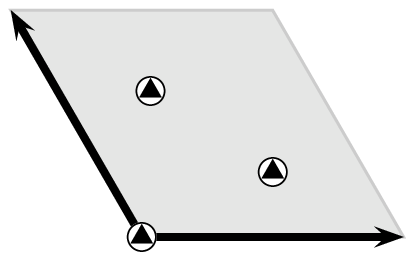}}
  &
  \CenterObject{\includegraphics[scale=0.8]{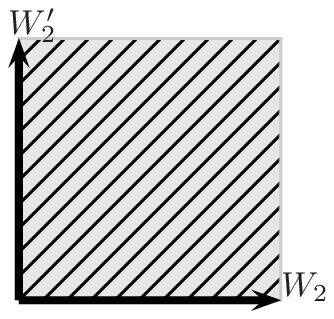}}
  \\[2cm]
  $\zb ~;~T_{3}$ &
  \CenterObject{\includegraphics[scale=0.8]{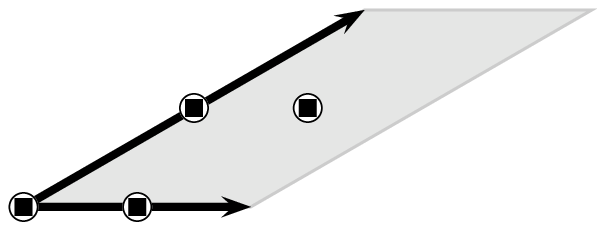}}
  &
  \CenterObject{\includegraphics[scale=0.8]{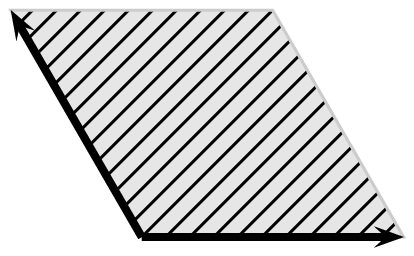}}
  &
  \CenterObject{\includegraphics[scale=0.8]{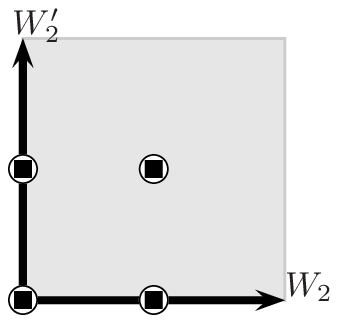}}
 \end{tabular}}
 \caption{
Fixed points and invariant planes (hatched) under the $\zc$ twist
and $\za,\zb$ subtwists, describing localization of different
twisted sectors. }
\label{fig:TableOfTori}
\end{figure}

For the $\G2 \times \SU3 \times \SO4$ lattice,
the action of the $\zc$ twist  is given by Eq.~\eqref{v6} with
\begin{equation}
v_6 = {1\over 6}(1,2,-3)\;.
\end{equation}
$z_1$, $z_2$ and $z_3$ are the coordinates of the $\G2$, $\SU3$ and
$\SO4$ $T^2$-tori, respectively. The $\zc$ twist $v_6$ has two subtwists,
\begin{equation}
\za :\; v_3 = 2 v_6 = {1\over 3}(1,2,-3)\;, \quad
\zb :\; v_2 = 3 v_6 = {1\over 2}(1,2,-3)\;.
\end{equation} 
An interesting feature of this orbifold is the occurrence of
invariant planes. Clearly, the $\za$ twist leaves the $\SO4$-plane invariant 
whereas the $\zb$ twist leaves the $\SU3$-plane invariant.
The corresponding fixed points and invariant planes are shown in
Fig.~\ref{fig:TableOfTori}. Our construction requires two Wilson lines in the
$\SO4$ plane, $W_2$ and $W_2'$, such that there are 3 independent gauge shift
vectors (\ref{3V6}) acting at different fixed points in this plane.

The rules of orbifold compactifications of the heterotic string have 
recently been reviewed in \cite{fnx04,krz042}. We are interested in the
states whose masses are small compared to the string scale $M_S$.
These states are described by fields
\begin{equation} 
\J_{r,s}(x;z_1,z_2,z_3)\;.
\end{equation}
Here $r$ labels the gauge quantum numbers and is given by 
\begin{equation}
 r\,= 
 \left\{\begin{array}{ll}
 p\quad& \text{for the untwisted sector}\;,\\
 p+k\,V_6\quad& \text{for the}\:k\text{-th twisted sector}\;, 
 \end{array}\right.
\end{equation}
where $p$ lies on the $\E8\times \E8$ root lattice  (\ref{root}) and   
we have absorbed the Wilson lines in the definition of
the $local$ twist $kV_6$.
Similarly, $s$ carries information about the spin,
\begin{equation}
 s\,= 
 \left\{\begin{array}{ll}
 q\quad& \text{for the untwisted sector}\;,\\
 q+k\,v_6\quad& \text{for the}\:k\text{-th twisted sector}\;, 
 \end{array}\right.
\end{equation}
where $q$ is an element of the $\SO8$ weight lattice
and $v_6^4=0$. 
In our convention, the last component of $q$ gives the 4D helicity.
For example, 4D vectors correspond to $q = (0,0,0,\pm 1)$,
4D scalars to  $q = (\pm 1,0,0,0)$ with all permutations of the 
first 3 entries, and fermions correspond to $q=\left(  \pm {1\over2},
\pm {1\over2},\pm {1\over2},\pm {1\over2}   \right)$ with an even number
of `+' signs\footnote{These $q$'s may have to be shifted by
an $\SO8$ root vector to satisfy masslessness conditions 
in twisted sectors.}.

The physical states are invariant under the action of the orbifold symmetry
group which consists of twists and translations. In our case 
only translations in the $\SO4$ plane 
have a non--trivial action on the gauge degrees of freedom,
due to the presence of Wilson lines.
Then the invariance conditions read\footnote{Here we omit 
string oscillator states.}
($\ell=1,..., 5$):
\begin{eqnarray}
\J_{r,s}(x;z_1,z_2,z_3) &=& e^{2\pi \I\, \ell (r\cdot V_6 - s\cdot v_6)}
\J_{r,s}\(x;e^{2\pi \I{\ell\over 6}}z_1, e^{2\pi \I {\ell\over 3}}z_2,
e^{-2\pi \I{\ell\over 2}}z_3\),\nonumber\\
\J_{r,s}(x;z_1,z_2,z_3) &=& e^{2\pi\I\, r\cdot W_2}
\J_{r,s}\(x;z_1,z_2,z_3+1\)\;,
\nonumber\\
\J_{r,s}(x;z_1,z_2,z_3) &=& e^{2\pi\I\, r\cdot W_2'}
\J_{r,s}\(x;z_1, z_2,z_3+\I \)\;,
\label{inv2}
\end{eqnarray}
where we have included the Wilson lines in the local shift vectors
$kV_6$. 
We note that here  two sources of symmetry breaking are present: local, due to
twisting, and non-local, due to the Wilson lines. In the
first case, symmetry breaking is restricted to the fixed points in the compact
space. Indeed, since  orbifold fixed points are invariant under twisting  (up to
a lattice vector), the first condition  can be satisfied only for certain $p$,
which indicates symmetry breaking at the fixed points. These sets of $p$'s are
generally different at different fixed points and only their intersection 
survives in 4D, since in this case the wave function can be constant in the
compactified dimensions leading to a massless state.  In the case of Wilson line
symmetry breaking, the second and third conditions  apply to all points in
the $\G2$ and $\SU3$ planes and the symmetry breaking is non-local. 

To define our string model, it is necessary to specify the action of the twist
on the second, `hidden', $\E8$. We find that the desired
symmetry breaking pattern and  the appearance of three $\boldsymbol{16}$-plets
at fixed points with unbroken $\SO{10}$ lead to
\begin{eqnarray} 
V_6 &=& \left({1\over 3},{1\over 3},{1\over 3},0,0,0,0,0\right)
\({1\over 6},{1\over 6},0,0,0,0,0,0\)\;,\\
W_2 &=& \left(0,{1\over 2},{1\over 2}, {1\over 2},{1\over 2},
                       0,0,0 \right)
\({1\over 2},{1\over 2},{1\over 2},0,0,0,0,{1\over 2}\)\;,\\
{W_2}' &=& \left( {1\over 4}, {1\over 4},- {1\over 4},{1\over 4},
-{1\over 4},-{3\over 4},-{3\over 4},-{3\over 4} \right)
\(0,{1\over 2},{1\over 2},{1\over 2},{1\over 2},0,0,0\)\;
\end{eqnarray}
in the orthonormal $\E8\times\E8$ basis. 
In string theory, these quantities must satisfy certain consistency
conditions (see \cite{fnx04} for a recent discussion).
First of all, $6V_6$ and $2W_2 , 2W_2'$ must be elements of 
the  $\E8\times\E8$ root lattice which is required by embedding of
the orbifold symmetry group (`space' group) in the gauge degrees 
of freedom. Second, modular invariance requires
\begin{equation}
6 \left[ (m V_6+ nW_2+n'W_2')^2 - m^2 v_6^2   \right] =0 {\rm ~ mod ~2},
~~m,n,n'=0,1 \;.
\end{equation}
Our choice of  the hidden sector components of $V_6,W_2,W_2'$ is 
strongly affected by these conditions.

We note that  $N=1$ supersymmetry in 4D requires
\begin{equation}
\sum_{i=1}^3 v_6^i =0 {\rm ~ mod ~1}\;,
\end{equation}
whereas $N=2$ would require, in addition, $v_6^i=0$ mod 1 for some $i$. 
In the former case, there is one gravitino satisfying $q\cdot v_6=0$ mod 1
whereas in the latter case there are two of them.

Finally, massless states in 4D must satisfy the following conditions: 
\begin{equation}
q^2=1 ~~,~~ p^2=2-2\tilde N
\end{equation}
for the untwisted sector, and
\begin{equation}
(q+kv_6)^2= c_{k} ~~,~~ \bigl(p+ kV_6 + n W_2 + n' W_2'\bigr)^2= c_{k,\tilde N}
\end{equation}
for the $k$-th twisted sector. Here $\tilde N$ is an oscillator
number and $c_{k},c_{k,\tilde N}$ are certain constants 
(see e.g. \cite{fnx04}). In our model, all states which transform
non--trivially under $\SU3_c \times \SU2_\mathrm{L}$ 
have $\tilde N=0$.

In this section we have described the necessary ingredients of our 
orbifold model. In the next section we compute the massless spectrum of
the model and discuss localization of various states.

\section{Massless spectrum of the model}

\begin{figure}[t]
 \centerline{\CenterObject{\includegraphics{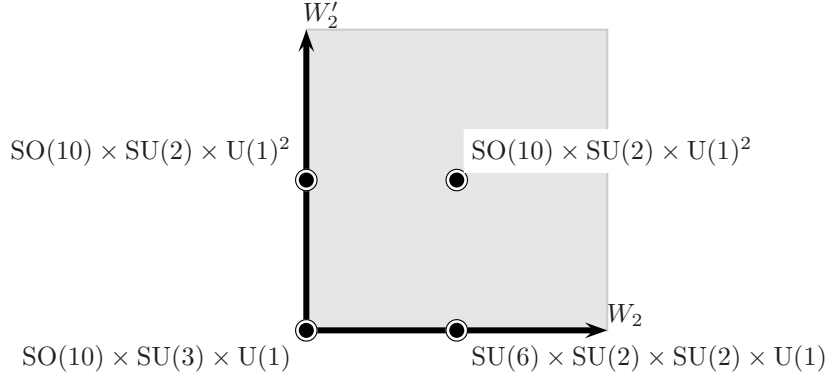}}}
 \caption{Local gauge symmetries in the $\SO4$-plane.}
 \label{fig:SO4Torus_GaugeGroups}
\end{figure}

First let us identify the gauge group in 4D.
For $N=1$ vector multiplets $q\cdot v_6 = 0$. Hence,
the surviving gauge group in 4D is given by the root vectors satisfying
\begin{equation}
p\cdot V_6,\ p\cdot W_2,\ p\cdot W_2'\ \in \z \;, \quad p^2 = 2\;.
\end{equation}
It is straightforward to verify that these roots together with the Cartan
generators form the Lie algebra of 
$$\SU3\times \SU2 \times \U1^5~~,$$
while the hidden sector  $\E8$ is broken to $\SU4 \times \SU4 \times \U1^2$. 
 This result can be understood by examining the enhanced 
gauge groups at the four orbifold fixed points in the $\SO4$-plane. 
These gauge groups are determined by
\begin{equation}
p\vspace{2cm}\cdot \Bigl(V_6+ n W_2 +n' W_2'\Bigr) \in \z \;,
\end{equation}
where $n,n'=\{ 0,1  \}$ specify  the fixed point in the $\SO4$ plane. Then,
omitting the hidden sector
the local gauge groups  are (Fig.~\ref{fig:SO4Torus_GaugeGroups}): 
\begin{eqnarray}
&& (n=0, n'= 0): ~\SO{10}\times \SU3\times \U1
\nonumber\\
&& (n=1, n'=0): ~\SU6\times \SU2\times \SU2\times
\mathrm{U}(1) \nonumber\\
&& (n=0, n'=1): ~\SO{10}\times \SU2\times 
\mathrm{U}(1)^2 \nonumber\\
&& (n=1, n'=1): ~\SO{10}\times \SU2\times \U1^2\;. 
\end{eqnarray}
These are precisely the groups discussed in the first section. Their
intersection yields the surviving group 
$\SU3\times \SU2\times \U1^5$. 

Let us now consider matter fields. These can be either
in the untwisted sector $U$ or in one of the  twisted
sectors $T_1\ldots T_5$. Below we analyze each of them separately.
Before we proceed, let us fix the chirality of the matter
fields to be $positive$\footnote{This is necessary to distinguish
matter fields from their CP conjugates.}, i.e. $q_4=+1/2$
for their fermionic components.

\subsection{$\boldsymbol{U}$ sector} 

For $N=1$ chiral multiplets in the untwisted sector we have 
$q\cdot v_6 = \pm 1/6$, $\pm 1/3$, $\pm 1/2$, and therefore
\begin{equation}
p\cdot V_6=\{ 1/6,1/3,1/2 \} ~{\rm mod}~ 1 \;;\quad
p\cdot W_2,\ p\cdot W_2'\ \in \z \;.
\end{equation}
The states represent bulk matter of the orbifold.  By choosing an appropriate
right--mover, these massless states can be made invariant under the 
$\mathbbm{Z}_6$ orbifold action and thus are present in the 4D spectrum. Apart
from $\SU3_c\times\SU2_\mathrm{L}$ singlets\footnote{We defer the analysis of
$\U1$ charges until a subsequent publication.}, the untwisted sector of our
model contains 
\begin{equation}
U=2\times(\boldsymbol{3},\boldsymbol{1})
 \oplus (\bar{\boldsymbol{3}},\boldsymbol{1}) 
 \oplus 5\times (\boldsymbol{1},\boldsymbol{2})
\end{equation}
in terms of the $\SU3_c\times \SU2_\mathrm{L}$ quantum numbers.
From the field-theory perspective, these fields correspond to the 
compact space
components of the $E_8$ gauge fields and their superpartners.

\subsection{$\boldsymbol{T_1 + T_5}$ sector}

These matter fields are located at the 12 orbifold fixed points (Fig.~\ref{fig:TableOfTori})
and satisfy 
\begin{equation}
\Bigl(p+V_6+ n W_2 +n' W_2'\Bigr)^2={25\over 18} \;.
\end{equation}
Since  Wilson lines are present only in the $\SO4$ plane,
only $\SO4$-plane projections of the fixed points   matter.
The $\mathrm{G}_2$ and $\SU3$ projections do not affect the 
local twist. They only lead to a multiplicity factor 3 due to the three 
identical $\SU3$ fixed points. 
Any massless state in the $T_1$ sector
survives the orbifold projection, i.e. is invariant under
the $\mathbbm{Z}_6$ action, and is therefore present in the 4D
spectrum.

The twisted matter fields located at a given fixed point appear in a 
representation of the local gauge group at this point. In our case, twisted
matter with  $\SU3_c\times\SU2_\mathrm{L}$ quantum numbers is
\begin{eqnarray}
&& (n=0, n'=0): ~3\times (\overline{\boldsymbol{16}},1) \nonumber\\
&& (n=1, n'=0): ~6\times (1,\boldsymbol{2},1)  \nonumber\\
&& (n=0, n'=1): ~ - \nonumber\\
&& (n=1, n'=1): ~ - 
\end{eqnarray}
It is convenient to keep the notation $(\overline{\boldsymbol{16}})$ of
$\SO{10}$ even though  the unbroken group in 4D is only $G_\mathrm{SM}$,
since it represents one complete generation of SM fermions including 
right--handed neutrinos. In terms of  $\SU3_c\times \SU2_\mathrm{L}$
quantum numbers we have
\begin{equation}
T_1 + T_5 = 3\times(\overline{\boldsymbol{16}}) \oplus 
6\times (\boldsymbol{1},\boldsymbol{2}) \;,
\end{equation}
where again we have omitted singlets.

\subsection{$\boldsymbol{T_2 +T_4}$ sector}

These states are localized at the fixed points in the  $\mathrm{G}_2$ and $\SU3$
planes, while being bulk states in the $\SO4$ plane
(Fig.~\ref{fig:TableOfTori}). If the $T_1$ sector corresponds to the string with
the boundary condition twisted by  $\Theta=\diag\left(e^{2\pi \I\, v_6^1},
e^{2\pi \I\, v_6^2},e^{2\pi \I\, v_6^3} \right)$, the $T_2$ sector corresponds
to the strings twisted by $\Theta^2$. Since $\Theta^2$ has a fixed plane, $T_2$
states are bulk states in this plane and localized states in the other two
planes. 

\begin{figure}[t]
 \centerline{\CenterObject{\includegraphics{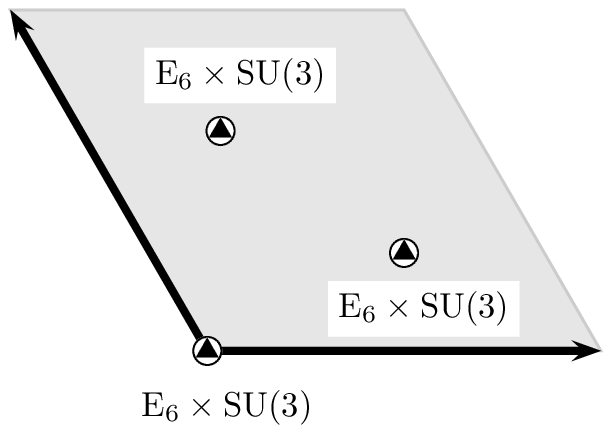}}}
 \caption{Local gauge symmetries in the $\SU3$-plane after
 the $\mathbbm{Z}_3$ subtwist.  } 
 \label{fig:SU3Torus_GaugeGroups_T2}
\end{figure}

The orbifold action on this sector is $\mathbbm{Z}_3$, and is given by
\begin{eqnarray}
 v_3=2v_6 ~~,~~ V_3 =2 V_6 ~\;.
\end{eqnarray}
Since there are no Wilson lines in the $\mathrm{G}_2$ and $\SU3$
planes, all fixed points are equivalent. The massless $N=1$ multiplets obey
\begin{equation}
\Bigl(p+V_3 \Bigr)^2={14\over 9} \;.
\end{equation}
Both the $\mathrm{G}_2$ and the $\SU3$ lattice have 3 fixed points 
under $\mathbbm{Z}_3$, so the multiplicity factor is 9. The local gauge 
groups at the fixed points are determined by
\begin{equation}
p \cdot V_3=0 \;.
\end{equation}
At each $\mathbbm{Z}_3$ fixed point, the unbroken gauge group and the twisted
sector matter fields are (cf.\ Fig.~\ref{fig:SU3Torus_GaugeGroups_T2})
\begin{equation}
\E6 \times \SU3 : \quad (\overline{\boldsymbol{27}},1)\;,
\end{equation}
plus $\SU3_c\times \SU2_\mathrm{L}$ singlets.

These states are subject to \emph{further} projection and  not all of them 
survive. Indeed, by construction they are only invariant under the
$\mathbbm{Z}_3$ action, but not under the full $\mathbbm{Z}_6$. Furthermore, 
the
$\mathbbm{Z}_3$ fixed points in the $\mathrm{G}_2$-plane are only fixed under
$\mathbbm{Z}_3$ and the $\mathbbm{Z}_6$ action transforms them into one
another. Physical states are formed out of their linear combinations which are
eigenstates of the $\mathbbm{Z}_6$ twist. 

The $\mathbbm{Z}_3$ invariance of a physical state requires
\begin{equation}
(q+v_3)\cdot v_3 = (p+V_3)\cdot V_3 {\rm ~mod~}1 \;,
\label{z3inv}
\end{equation}
where 
$q+v_3$ is
the shifted $\SO8$ momentum and $p+V_3$ is the shifted $\mathrm{E}_8
\times \mathrm{E}_8$ momentum. This is satisfied automatically as long as  the
gauge embedding of the twist and the Wilson lines obey modular invariance 
(yet it may require shifts by a lattice vector). A non-trivial $\mathbbm{Z}_2$
invariance condition is
\begin{equation}
(q+v_3)\cdot v_2 = (p+V_3)\cdot V_2  ~+~ \gamma   {\rm ~mod~}1  \;, \quad
p\cdot W_2,\ p\cdot W_2'\ \in \z \;,
\label{z2inv}
\end{equation}
where 
\begin{equation}
v_2=3v_6 ~~,~~ V_2 =3 V_6 \;. 
\end{equation}
The extra term
$\gamma=(0,0,1/2)$ appears due to the ``mixing'' of the fixed 
points \cite{Kobayashi:1990mc},\cite{krz041}. There are
three  combinations of the $\mathbbm{Z}_3$ fixed points which are eigenstates
of $\mathbbm{Z}_6$ with eigenvalues  $e^{2\pi \I \gamma}$. 

An important note is in order. The $\SO8$ lattice momentum $q$
is found via the masslessness condition for the right--movers,
\begin{equation}
(q+v_3)^2={5\over 9}\;.
\end{equation}
Since $v_3$ has a fixed plane, there are always two sets of solutions,
with opposite chiralities. Both of them survive 
the projection (\ref{z3inv}), which leads to $N=2$ hypermultiplets.
The conditions (\ref{z2inv}) break the symmetry between the
two chiralities and one obtains  $N=1$ chiral multiplets.

As a result, $9\times (\overline{\boldsymbol{27}})$ $N$=$2$ hypermultiplets
produce the following $N$=$1$ multiplets with 
$\SU3_c\times \SU2_\mathrm{L}$ quantum numbers:
\begin{equation}
T_2 +T_4 = 3\times (\boldsymbol{3},\boldsymbol{1}) 
\oplus 6\times (\bar{\boldsymbol{3}},\boldsymbol{1}) 
\oplus 9\times(\boldsymbol{1},\boldsymbol{2}) \;.
\end{equation}

\subsection{$\boldsymbol{T_3}$ sector}

These states are localized at the $\mathbbm{Z}_2$ fixed points in the
$\mathrm{G}_2$ and $\SO4$ planes and are bulk states in the
$\SU3$ plane (Fig.~\ref{fig:TableOfTori}). 
They correspond to strings twisted by $\Theta^3$. The
massless $T_3$ states satisfy 
\begin{equation}
\Bigl(p+V_2 + n W_2 +n' W_2' \Bigr)^2={3\over 2} \;,
\end{equation}
and the local gauge groups at the fixed points are determined by
\begin{equation}
p\cdot \Bigl(V_2+ n W_2 +n' W_2'\Bigr)=0 \;.
\end{equation}
The result for gauge groups and matter multiplets reads
\begin{eqnarray}
&& (n=0, n'= 0): ~\SO{16} ~,~ 8\times(\boldsymbol{16}) \nonumber\\
&& (n=1, n'=0): ~\SO{16} ~,~ 8\times(\boldsymbol{16})   \nonumber\\
&& (n=0, n'=1): ~\E7\times \SU2~,~ - \nonumber\\
&& (n=1, n'=1): ~\E7\times \SU2~,~- \quad.
\end{eqnarray}
As usual we have omitted  $\SU3_c\times \SU2_\mathrm{L}$   singlets and 
included a multiplicity  factor 4 from the $\mathrm{G}_2$-plane fixed points.
These states are located at the $\mathbbm{Z}_2$ fixed points which are mixed by
the action of the full $\mathbbm{Z}_6$ twist. Again, one has to form linear
combinations of the states transforming  covariantly under $\mathbbm{Z}_6$.

\begin{figure}[t]
 \centerline{\CenterObject{\includegraphics{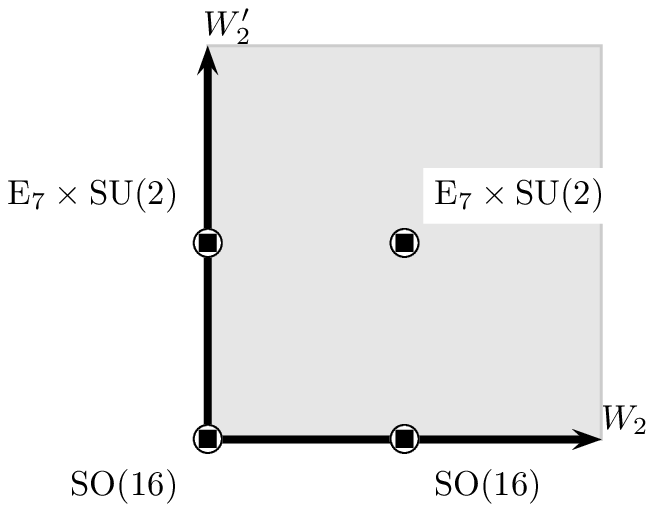}}}
 \caption{Local gauge symmetries in the $\SO4$-plane
 after the $\mathbbm{Z}_2$ subtwist. } 
 \label{fig:SU3Torus_GaugeGroups_T3}
\end{figure}

The matter states are, as before, subject to projection
conditions. The $\mathbbm{Z}_2$ condition for the relevant states is 
\begin{equation}
(q+v_2)\cdot v_2 = (p+V_2 +n W_2)\cdot ( V_2+ nW_2)  {\rm ~mod~}1 \;. 
\end{equation}
With a proper redefinition of $W_2$ by a lattice vector shift,
this condition is satisfied by all states with  both chiralities, i.e.
both solutions $q$ of the equation for massless right--movers, 
\begin{equation}
(q+v_2)^2={1\over 2}\;.
\end{equation}
Therefore, these states form $N=2$ hypermultiplets.
The further $\mathbbm{Z}_3$ projection reads 
\begin{equation}
 (q+v_2)\cdot v_3 = (p+V_2+nW_2)\cdot V_3  ~+ ~\gamma   {\rm ~mod~}1  \;,
\end{equation}
where now $\gamma=(0,0,1/3,-1/3)$. 
The four $\mathbbm{Z}_2$ fixed points in the $\mathrm{G}_2$-plane lead to
four eigenstates under $\mathbbm{Z}_3$ with eigenvalues $e^{2\pi \I \gamma}$.
The above  condition projects out some of the states. 
The surviving  $N$=1 multiplets
with $\SU3_c\times \SU2_\mathrm{L}$ quantum numbers are
\begin{equation}
T_3 = 7\times (\boldsymbol{3},\boldsymbol{1}) \oplus
 5\times (\bar{\boldsymbol{3}},\boldsymbol{1}) 
 \oplus 10\times(\boldsymbol{1},\boldsymbol{2}) \;.
\end{equation}

\subsection{Summary of the massless spectrum}

Combining all matter multiplets from the untwisted and the five twisted
sectors we finally obtain 
\begin{eqnarray}
M &=& U + T_1 + T_2 + T_3 + T_4 + T_5 \nonumber\\
 &=& 3\times ({\overline{\boldsymbol{16}}}) 
\oplus 12\times ( \boldsymbol{3},\boldsymbol{1}) 
\oplus 12\times (\bar{\boldsymbol{3}},\boldsymbol{1}) 
\oplus 30\times (\boldsymbol{1},\boldsymbol{2})\;, 
\end{eqnarray}
plus $\SU3_c\times \SU2_\mathrm{L}$ singlets. Note that in addition to three SM generations contained
in the three ${\bf 16}$-plets we have only \emph{vector--like} matter.
This result is partly dictated by the requirement of anomaly cancellations.
Vector--like fields can attain large masses and decouple from
the low energy theory. A detailed analysis of this issue, including 
$\mathrm{U(1)}$ factors, will be presented in a subsequent publication.

\section{Intermediate GUTs}

So far we have made no assumption on the size of the compact dimensions.
These  are usually assumed to be given by the string scale, $R_i \sim 1/M_S$.
However, this is not necessarily the case and, furthermore, 
unification of the gauge couplings favours anisotropic compactifications 
where some of the radii are significantly larger than the others 
\cite{wit96,ht04}. 
In this case one encounters a higher--dimensional GUT at
an intermediate energy scale. Indeed, the Kaluza--Klein modes
associated with a large dimension of radius $R$  become light
and are excited at energy scales above $1/R \ll M_S$. 
At these energy scales we obtain an effective higher--dimensional field theory
with enhanced symmetry in the bulk.

In our model  there are four independent radii: two are associated with
the $\G2$ and $\SU3$ planes, respectively, and the other two 
are associated with the two independent directions in the $\SO4$-plane.
Any of these radii can in principle be large leading to a distinct
GUT model.
\begin{figure}[t]
 \centerline{\CenterObject{\includegraphics{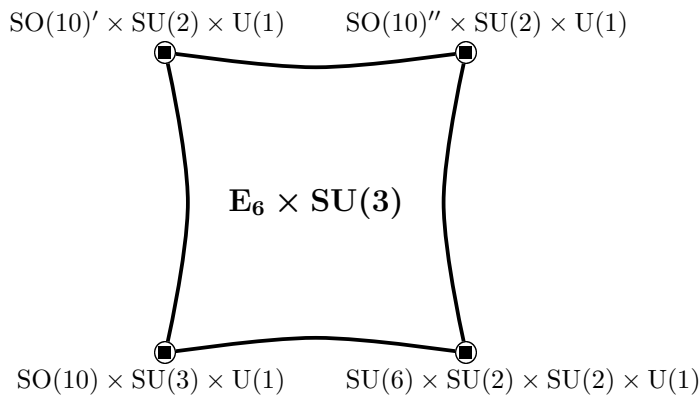}}}
\caption{6D $\E6\times \SU3$ orbifold GUT for a large
 compactification radius of the $\SO4$-plane.} 
 \label{fig:SO4Pillow_T2}
\end{figure}

The bulk gauge group and the amount of supersymmetry are found
via a subset of the invariance conditions (\ref{inv2}).
Consider a subspace ${\cal S}$  of the 6D compact space
with large compactification radii. This subspace is left invariant
under the action of some elements  of the orbifold space group, i.e.
a subset of twists and translations ${\cal G }$. 
The bulk gauge multiplet  in ${\cal S}$
is a subset of the $N=4$ $\E8 \times \E8$ gauge multiplet
which is invariant under the action of ${\cal G }$,
i.e. a subset of conditions (\ref{inv2}) restricted to ${\cal G }$.

Consider first  the case with two large compact dimensions, 
for instance  those associated with the $\SO4$-plane. 
The $\SO4$-plane  is invariant under the $\za$ subtwist as well
as translations by a lattice vector  
in the $\G2$ and $\SU3$ planes. The latter
do not lead to non--trivial projection conditions
since there are no Wilson lines in these planes, 
while the former leads to gauge symmetry and supersymmetry breaking.
The light gauge states are  described by fields
which are constant with respect to  $z_1$ and $z_2$. 
Invariance under $\za$ requires (see Eq.(\ref{inv2}) with $\ell=2$)
\begin{eqnarray}
\J_{p,q}(x;z_3) &=& e^{2\pi\I\,  (p\cdot V_3 - q\cdot v_3)}
\J_{p,q}\(x;z_3\)\;.
\label{so4gut}
\end{eqnarray}
Gauge multiplets satisfy $q\cdot v_3=0$ which has two sets
of solutions for $q$ corresponding to $N=2$ supersymmetry.
Then the condition $p\cdot V_3=0$ breaks $\E8$
to $\E6\times \SU3$. At the four fixed points of the
$\SO4$-plane symmetry  is broken further to the four subgroups
discussed in Sect.~3. Altogether, we obtain a 6D $\E6\times \SU3$ orbifold GUT with  the distribution of gauge symmetries in the fundamental region of the
orbifold shown in Fig.~\ref{fig:SO4Pillow_T2}.
Similarly, untwisted matter satisfies (\ref{so4gut}) with 
$q\cdot v_3=\pm 1/3 $. We note that all
three SM generations live at the origin in Fig.~\ref{fig:SO4Pillow_T2}.

\begin{figure}[t]
 \centerline{\CenterObject{\includegraphics{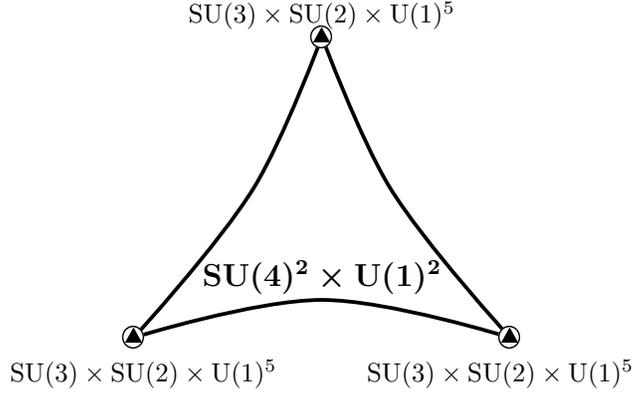}}}
 \caption{6D $\SU4\times \SU4\times \U1^2$ orbifold GUT for a large 
  compactification radius of the $\SU3$-plane.}
 \label{fig:SU3Pillow_T2}
\end{figure}
A similar analysis can be carried out for the $\SU3$-plane, 
which is invariant under the $\zb$ subtwist and
lattice translations in the $\G2$ and $\SO4$ planes.
In this case, there are also non--trivial projection conditions
due to the Wilson lines,
\begin{eqnarray}
\J_{p,q}(x;z_2) &=& e^{2\pi \I\,  (p\cdot V_2 - q\cdot v_2)}
\J_{p,q}\(x;z_2 \),\nonumber\\
\J_{p,q}(x;z_2) &=& e^{2\pi\I\, p\cdot W_2}
\J_{p,q}\(x;z_2\)\;,
\nonumber\\
\J_{p,q}(x;z_2) &=& e^{2\pi\I\, p\cdot W_2'}
\J_{p,q}\(x;z_2\)\;.
\end{eqnarray}
This breaks $N=4$ $\E8$ to $N=2$ $\SU4\times \SU4\times
\U1^2$ in the bulk (Fig.~\ref{fig:SU3Pillow_T2}). 
At the fixed points, 
the symmetry is broken further by the $\za$ twist 
leaving only the standard model gauge group (up to $\U1$'s).
Each of the three fixed points carries one generation
of the standard model matter.
\begin{figure}[t]
 \centerline{\CenterObject{\includegraphics{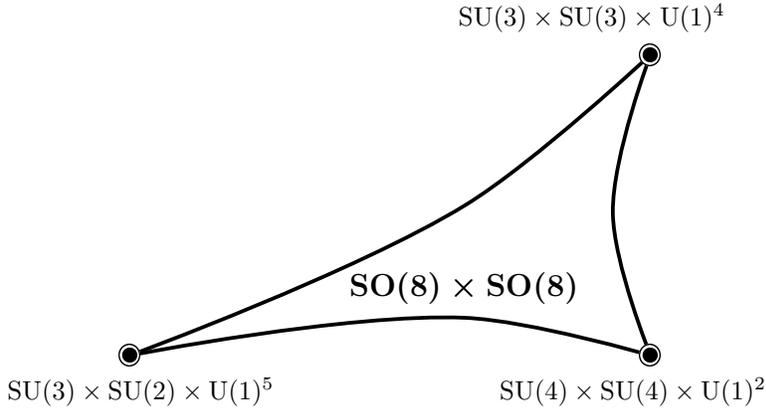}}}
 \caption{6D $\SO8\times \SO8$
 orbifold GUT for a large $\G2$-plane compactification radius.}
 \label{fig:G2Pillow_T2}
\end{figure}

A different picture arises when 
the $\G2$-plane compactification radius is large.
The $\G2$-plane is not invariant under any of the twists,
thus there is no projection condition due to twisting.
The only non--trivial projection conditions are due to the Wilson
lines,
\begin{eqnarray}
\J_{p,q}(x;z_1) &=& e^{2\pi\I\, p\cdot W_2}
\J_{p,q}\(x;z_1\)\;,
\nonumber\\
\J_{p,q}(x;z_1) &=& e^{2\pi\I\, p\cdot W_2'}
\J_{p,q}\(x;z_1\)\;.
\end{eqnarray}
Thus we have $N=4$ supersymmetry and the gauge group is
$\SO8\times \SO8$. Three generations of the standard model are
localized at the origin where the $\zc$ twist
breaks the symmetry to the standard model gauge group
(Fig.~\ref{fig:G2Pillow_T2}).

\begin{table}[!ht]
\begin{center}
\begin{tabular}{|r|ccc|l|l|}
\hline
 & \multicolumn{3}{|c|}{plane} & 
& \\
dim. & $\mathrm{G}_2$ & $\SU3$ & $\SO4$ & conditions
& SUSY, bulk groups\\
\hline
10&
\CenterObject{\includegraphics[scale=0.3]{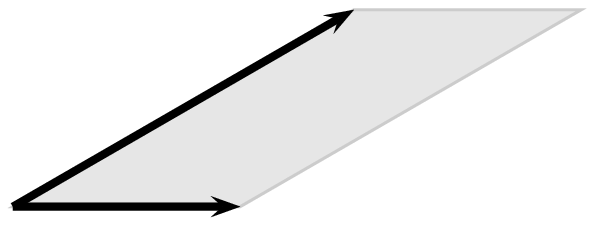}}
& 
\CenterObject{\includegraphics[scale=0.3]{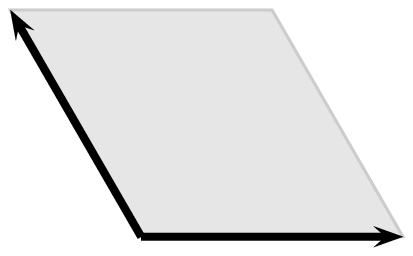}}
& 
\CenterObject{\includegraphics[scale=0.3]{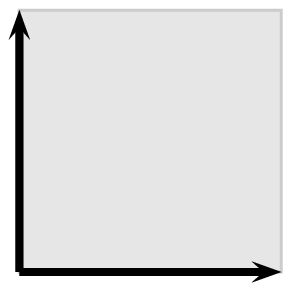}}
& -- & $N=4$, $\mathrm{E}_8$ \\[2mm]
\hline
9&
\CenterObject{\includegraphics[scale=0.3]{G2Torus.eps}}
& 
\CenterObject{\includegraphics[scale=0.3]{SU3Torus.eps}}
& 
\CenterObject{\includegraphics[scale=0.3]{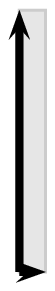}}
& $p\cdot W_2\in\mathbbm{Z}$ & $N=4$, $\SO{16}$ \\[2mm]
9&
\CenterObject{\includegraphics[scale=0.3]{G2Torus.eps}}
& 
\CenterObject{\includegraphics[scale=0.3]{SU3Torus.eps}}
& 
\CenterObject{\includegraphics[scale=0.3]{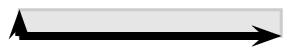}}
& $p\cdot W_2'\in\mathbbm{Z}$ & $N=4$, $\SO{16}$ \\[2mm]
\hline
8&
\CenterObject{\Huge$\bullet$}
& 
\CenterObject{\includegraphics[scale=0.3]{SU3Torus.eps}}
& 
\CenterObject{\includegraphics[scale=0.3]{SO4Torus.eps}}
& -- &  $N=4$, $\mathrm{E}_8$\\[2mm]
8&
\CenterObject{\includegraphics[scale=0.3]{G2Torus.eps}}
& 
\CenterObject{\Huge$\bullet$}
& 
\CenterObject{\includegraphics[scale=0.3]{SO4Torus.eps}}
& -- &  $N=4$, $\mathrm{E}_8$ \\[2mm]
8&
\CenterObject{\includegraphics[scale=0.3]{G2Torus.eps}}
& 
\CenterObject{\includegraphics[scale=0.3]{SU3Torus.eps}}
& 
\CenterObject{\Huge$\bullet$}
&
$p\cdot W_2,\, p\cdot W_2'\in\mathbbm{Z}$ 
& $N=4$, $\SO8\times\SO8$ 
\\[2mm]
\hline
7&
\CenterObject{\Huge$\bullet$}
& 
\CenterObject{\includegraphics[scale=0.3]{SU3Torus.eps}}
& 
\CenterObject{\includegraphics[scale=0.3]{SO4TorusHigh.eps}}
& $p\cdot W_2\in\mathbbm{Z}$ & $N=4$, $\SO{16}$ \\[2mm]
7&
\CenterObject{\Huge$\bullet$}
& 
\CenterObject{\includegraphics[scale=0.3]{SU3Torus.eps}}
& 
\CenterObject{\includegraphics[scale=0.3]{SO4TorusLong.eps}}
& $p\cdot W_2'\in\mathbbm{Z}$ & $N=4$, $\SO{16}$ \\[2mm]
7&
\CenterObject{\includegraphics[scale=0.3]{G2Torus.eps}}
& 
\CenterObject{\Huge$\bullet$}
& 
\CenterObject{\includegraphics[scale=0.3]{SO4TorusHigh.eps}}
& $p\cdot W_2\in\mathbbm{Z}$ & $N=4$, $\SO{16}$ \\[2mm]
7&
\CenterObject{\includegraphics[scale=0.3]{G2Torus.eps}}
& 
\CenterObject{\Huge$\bullet$}
& 
\CenterObject{\includegraphics[scale=0.3]{SO4TorusLong.eps}}
& $p\cdot W_2'\in\mathbbm{Z}$ & $N=4$, $\SO{16}$ \\[2mm]
\hline
6&
\CenterObject{\Huge$\bullet$}
& 
\CenterObject{\Huge$\bullet$}
& 
\CenterObject{\includegraphics[scale=0.3]{SO4Torus.eps}}
& $p\cdot 2V_6\in\mathbbm{Z}$ &  $N=2$, $\E6\times\SU3$\\[2mm]
6&
\CenterObject{\Huge$\bullet$}
& 
\CenterObject{\includegraphics[scale=0.3]{SU3Torus.eps}}
& 
\CenterObject{\Huge$\bullet$}
& $p\cdot 3V_6,\, p\cdot W_2,\, p\cdot W_2'\in\mathbbm{Z}$ 
&  $N=2$, $\SU4\times\SU4$ \\[2mm]
6&
\CenterObject{\includegraphics[scale=0.3]{G2Torus.eps}}
& 
\CenterObject{\Huge$\bullet$}
& 
\CenterObject{\Huge$\bullet$}
& 
$p\cdot W_2,\, p\cdot W_2'\in\mathbbm{Z}$ 
& $N=4$, $\SO{8}\times\SO{8}$ \\[2mm]
\hline
5&
\CenterObject{\Huge$\bullet$}
& 
\CenterObject{\Huge$\bullet$}
& 
\CenterObject{\includegraphics[scale=0.3]{SO4TorusHigh.eps}}
& 
$p\cdot 2V_6,\,p\cdot W_2\in\mathbbm{Z}$ 
& $N=2$, $\SU6\times\SU2^2$ \\[2mm]
5&
\CenterObject{\Huge$\bullet$}
& 
\CenterObject{\Huge$\bullet$}
& 
\CenterObject{\includegraphics[scale=0.3]{SO4TorusLong.eps}}
& 
$p\cdot 2V_6,\, p\cdot W_2'\in\mathbbm{Z}$ 
& $N=2$, $\SU6\times\SU2^2$ \\[2mm]
\hline
4&
\CenterObject{\Huge$\bullet$}
& 
\CenterObject{\Huge$\bullet$}
& 
\CenterObject{\Huge$\bullet$}
& 
$p\cdot V_6,\, p\cdot W_2,\, p\cdot W_2'\in\mathbbm{Z}$ 
& 
$\hspace*{-0.2cm}\begin{array}{l}
 N=1,~\SU3\times\SU2\\
\sim G_\mathrm{SM}\end{array}$ \\[2mm]
\hline
\end{tabular}
\end{center}
\caption{Survey of the various orbifold GUTs in different
dimensions. The bullet indicates small compact dimensions. 
$\U1$ factors are
omitted. }
\end{table}

In principle, there is nothing special about six dimensions, and the same
analysis can be carried out for five, seven, eight, nine and ten dimensions.
The results are summarized in  Table 1. A variety of orbifold GUTs appears,
with gauge groups ranging from $\E8$ to $\SU4\times \SU4\times \U1^2$.
These GUTs represent different points in moduli space. Values of the
corresponding T-moduli determine the compactification radii.

It is remarkable that all these GUT models in various dimensions 
are consistent with gauge coupling 
unification\footnote{Here we only consider  running of the
gauge couplings in the bulk. An analysis of localized contributions
will be presented elsewhere.}. 
This is true  even though in some cases $\SU3_c$ and $\SU2_\mathrm{L}$ 
are contained in different simple factors, i.e. 
$\SO8\times \SO8$ or $\SU4 \times \SU4$.
The beta functions for both $\SO8$'s or $\SU4$'s are the same.
In  the former case this is enforced by $N=4$ supersymmetry,
while in the latter case the two $\SU4$'s have identical 
bulk matter content, $2\times(\boldsymbol{6,1})+ 16\times (\boldsymbol{4,1})$
$N=2$ multiplets. In all other cases $\SU3_c\times \SU2_\mathrm{L}$ 
is contained in a
simple factor such that unification of the gauge couplings in the bulk
is automatic.

On the other hand, different  GUTs 
differ in the value of the gauge coupling at the unification scale, 
since the power law running depends on the number of extra dimensions and
the bulk gauge group. 
Realization of some of the GUTs may  require nonperturbative
string coupling \cite{wit96,ht04}.
Different models also lead to  different Yukawa couplings
which depend on the compactification radii.
These phenomenological aspects are similar to those of orbifold GUTs 
\cite{abc03} and will be discussed elsewhere.

\noindent
\section{Summary}

We have presented a $\zc$ heterotic orbifold
model leading to the standard model spectrum and
additional vector--like matter in four dimensions.
Standard model generations appear as $\boldsymbol{16}$-plets
of $\SO{10}$. They are localized at
different fixed points in the compact space with local
$\SO{10} \times \SU3 \times \U1$ symmetry.

If some of the compactification radii are significantly
larger than the others, we recover various higher--dimensional GUTs
as an intermediate step at energies below $M_S$.
These GUTs have the same 4D massless spectrum and the same
ultraviolet completion, but represent different points in moduli space.
All of them are consistent with gauge coupling unification, yet differ 
in other phenomenological aspects.  \\

\noindent
{\bf Acknowledgements.} 
We would like to thank S.~F\"orste, A.~Hebecker, T. Kobayashi,  
H.-P.~Nilles, M.~Trapletti, P.~K.~S.~Vaudrevange and A.~Wingerter
for discussions.
One of us (M.R.) would like to thank the Aspen Center for Physics for support.
This work was partially supported by the EU 6th Framework Program
MRTN-CT-2004-503369 ``Quest for Unification'' and MRTN-CT-2004-005104
``ForcesUniverse''.



\end{document}